# One-way heat transfer in deep-subwavelength thermophotonics

Shuihua Yang[1,#], Chen Jianfeng[1,#], Guoqiang Xu[1], Jiaxin Li[1], Xianghong Kong[1], and Cheng-Wei Qiu[1,*]

[1]*Department of Electrical and Computer Engineering, National University of Singapore, Kent Ridge 117583, Republic of Singapore.*

[#]These authors contributed equally to this work.

[*]Correspondence to: chengwei.qiu@nus.edu.sg

**Abstract**

Nonreciprocal thermophotonics, by breaking Lorentz reciprocity, exceeds current theoretical efficiency limits, unlocking opportunities to energy devices and thermal management. However, energy transfer in current systems is highly defect-sensitive. This sensitivity is further amplified at deep subwavelength scales by inevitable multi-source interactions, interface wrinkles, and manufacturing tolerances, making precise control of thermal photons increasingly challenging. Here, we demonstrate a topological one-way heat transport in a deep-subwavelength thermophotonic lattice. This one-way heat flow, driven by global resonances, is strongly localized at the geometric boundaries and exhibits exceptional robustness against imperfections and disorder, achieving nearly five orders of radiative enhancement. Our findings offer a blueprint for developing robust thermal systems capable of withstanding strong perturbations.

**Introduction**

Fresnel predicted that momentum interactions between photons and moving media would lead to a dragging effect on light [1]. This phenomenon was later experimentally verified by Fizeau [2], who demonstrated that light propagates at different speeds when moving along or against the water flow (Fig. 1a). Similar effects have been observed in various moving or spatiotemporal media [3-5].

This groundbreaking discovery sparked interest in how electron flows can drag polaritons in solid-state systems (Fig. 1b). For instance, the plasmonic Fizeau effect has been investigated in graphene under an external bias voltage, where time-reversal symmetry breaking originates from the nonlinear electrodynamics of Dirac electrons in the material [6-8]. These nonreciprocal surface modes have been employed to develop radiative thermal diodes and enhance near-field thermal radiation at subwavelength scales [9, 10]. However, surface-mode-mediated thermal radiation is highly sensitive to external factors, including Joule heating from the applied voltage, higher-order Fizeau effects from edge reflections, and scattering losses due to material wrinkles, all of which degrade performance. In practical applications, this sensitivity, combined with the stringent demands for device quality, causes energy transfer paths vulnerable to environmental perturbations and local defects, severely limiting the development of robust thermal devices.

Integrating topological concepts into photonic systems provides promising strategy for robust energy manipulation, with topologically protected edge states being the most notable feature [11-15]. While research in topological thermophotonics has explored concepts such as Zak phase, higher-order topology [16, 17], quasicrystals [18], and spin-Hall-like phase [19], these states remain constrained by strict symmetry. Imperfections in structure and transmission path can lead to strong scattering or even the disappearance of the topological states [20, 21]. For instance, while Kramers' degeneracy prevents scattering between electrons of different spins in fermionic systems, this fundamental protection is absent for electromagnetic waves and light. As a result, defects that induce pseudo-spin flipping can cause strong backscattering, inevitably degrading the quality of the radiative output. This challenge is especially pronounced at deep-subwavelength scales, where inevitable multi-source interactions, interface wrinkles, and manufacturing tolerances make the manipulation of thermal photons particularly challenging.

In this Letter, we present a topological thermophotonic lattice that achieves robust one-way heat transfer at deep subwavelength scales (Fig. 1c). This one-way heat flow propagates along the interface without backscattering. Intriguingly, due to global topological protection, the energy flow maintains strong unidirectionality and high transmission efficiency even when encountering sharp bends or scattering obstacles. Additionally, global resonances induced by multi-source interactions strongly localize the heat flux at the boundaries, further enhancing the thermophotonic response. Quantitative calculations show that this synergistic combination of one-way propagation, strong localization, and robustness increases radiative heat transfer by nearly five orders of magnitude.

**Results and Discussion**

*Circular heat flux.* An essential prerequisite for realizing nonreciprocal thermophotonic systems is breaking time-reversal symmetry. Inspired by the Corbino effect in electronics—where a magnetic field perpendicular to a metal disk plane causes tangential currents to merge with radial currents, forming "circular" currents—a magnetic field can similarly break time-reversal symmetry in thermophotonic systems, leading to heat flux deflection [22, 23]. Here, we introduce a magnetic-free Weyl semimetal, characterized by two Weyl nodes with opposite chirality, separated by a tunable wavevector 2**b** at the same energy level [24, 25]. In this case, the momentum separation of the Weyl nodes functions similarly to the axial vector of an internal magnetic field, enabling time-reversal symmetry breaking in its perpendicular plane without requiring the external bias. This ideal magnetic Weyl semimetal phase has been experimentally realized in materials such as $K_2Mn_3(AsO_4)_3$ and $EuCd_2As_2$ under external magnetic fields [26, 27]. The thermophotonic Corbino effect can be compactly described by axion electrodynamics, which incorporates an axion term into the electromagnetic Lagrangian density:

$$\mathcal{L}_\theta = \frac{e^2}{2\pi\epsilon_0 \hbar c} \sqrt{\frac{\epsilon_0}{\mu_0}} \frac{(2\mathbf{b} \cdot \mathbf{r})}{2\pi} \mathbf{E} \cdot \mathbf{B} \qquad (1)$$

Here, $e$ is the elementary charge, $\hbar$ is the reduced Planck constant, $\epsilon_0$ is the permittivity of vacuum, $\mu_0$ is the permeability of vacuum, **E** is the electric field, and **B** is the magnetic flux density. The axion term changes the electromagnetic response of the Weyl semimetal, one direct manifestation being the generation of circular heat flux. It is noted that the thermophotonic Corbino effect can also be extended to other magneto-optical materials. As shown in Fig. 2a, the deflection of heat flux can be qualitatively described by the off-diagonal polarization component of the nanoparticles $\hat{\alpha}_{12}$ (see Supplementary Note I). It is seen that the direction of circular heat flux can be directly altered by the Weyl node wavevector, while degenerate Weyl nodes only support omnidirectional thermal emission (Fig. S2). Intriguingly, changing the excitation frequency can also reverse the direction of the circular heat flux (inset of Fig. 2), unlike the magnetic field dependence seen in traditional magneto-optical materials. By weighting the near-field ($1/(k_0 r)^3$) and far-field ($1/k_0 r$) contributions, we confirm that near-field effects dominate within a 500 nm range (Fig. 2).

*Thermophotonic crystal and bulk band structure.* We propose utilizing a honeycomb thermophotonic lattice to demonstrate tunable one-way radiative heat transfer. Each unicell consists of two nanoparticles belonging to the sublattices A and B, which exhibit π-rotation symmetry (Fig. 1c). The nanoparticles have a radius of $R = 120$ nm, and the distance between adjacent A and B sites is $d = 360$ nm. Given that the thermal wavelength is much larger than the lattice constant ($\sqrt{3}d$), long-range coupling becomes significant, making traditional tight-binding theory inadequate to describe such complex deep-subwavelength systems. Based on many-body radiative heat transfer theory [28-30], the effective Hamiltonian for the 2D thermophotonic lattice is expressed as:

$$\mathcal{H}^{\text{OP/IP}} = \begin{bmatrix} \left[k_0^2 \sum_{j \neq i} \hat{\alpha} \mathbb{G}_{A_i A_j}^{E,\text{OP/IP}} e^{i\mathbf{k}\cdot(\mathbf{r}_{Aj}-\mathbf{r}_{Ai})}\right] & \left[k_0^2 \sum_j \hat{\alpha} \mathbb{G}_{A_i B_j}^{E,\text{OP/IP}} e^{i\mathbf{k}\cdot(\mathbf{r}_{Bj}-\mathbf{r}_{Ai})}\right] \\ \left[k_0^2 \sum_j \hat{\alpha} \mathbb{G}_{B_i A_j}^{E,\text{OP/IP}} e^{i\mathbf{k}\cdot(\mathbf{r}_{Bj}-\mathbf{r}_{Ai})}\right] & \left[k_0^2 \sum_{j \neq i} \hat{\alpha} \mathbb{G}_{B_i B_j}^{E,\text{OP/IP}} e^{i\mathbf{k}\cdot(\mathbf{r}_{Bj}-\mathbf{r}_{Bi})}\right] \end{bmatrix} \quad (2)$$

where $\mathbb{G}$ represents the Green's tensors. The symmetry of Green's tensor ensures that the in-plane (IP) and out-of-plane (OP) modes are decoupled, enabling their band structures to be evaluated separately. In this work, we orient the crystal of the Weyl semimetal such that **b** is aligned along the OP direction. Thus, we restrict the Green's function to the *x–y* subspace. Figures 2b and 2c depict the topological phase transition and frequency shift induced by nonzero Weyl node separation. The nontrivial bandgap, driven by thermophotonic Corbino effect and many-body interactions, can be quantitatively evaluated through topological charges (see Supplementary Note III).

To elucidate the properties of one-way edge states, we consider the projective band structure of a 1D thermophotonic superlattice. The system is configured with open boundary conditions (OBC) along the longitudinal direction and periodic boundary conditions (PBC) along the transverse diraction, featuring a zigzag boundary. In the nontrivial case ($b$ = 0.8 nm$^{-1}$), two in-gap dispersion curves intersect at $k_x = \pi/a$, as shown in Fig. 2d. Generally, the slope of these dispersion curves indicates the propagation direction of the corresponding eigenmodes. Thus, the in-gap eigenstates will propagate in the positive (right) and negative (left) directions of the geometric edge (right panel of Fig. 2d). Conversely, in a trivial lattice ($b$ = 0 nm$^{-1}$), the bandgap closes, and the edge states degenerate, resulting in the loss of chirality and directional transport (see Supplementary Fig. S3c).

We also construct a hexagonal geometry to compare the heat transfer characteristics in nontrivial and trivial lattices (Supplementary Fig. S4). In the nontrivial lattice, the radiative heat flux is confined along the edges and propagates counterclockwise, thereby functioning similarly to a thermal circulator. Notably, the one-way energy flow persists even at bends, validating the presence of nonreciprocal edge modes with true backscattering immunity. In contrast, the trivial lattice exhibits isotropic energy flow, with propagation strength significantly reduced by several orders of magnitude.

*Topological nonreciprocity vs. Classical nonreciprocity.* Nonreciprocal systems can be categorized into topological and classical based on their bulk topology. In topologically nonreciprocal systems, a nonzero Weyl node separation is applied globally, endowing the system with topological protection (left panel of Fig. 3a). Additionally, the presence of global resonances within the system significantly enhances radiative heat transfer. The topological bandgap further localizes this resonantly amplified energy at the geometric boundaries, ensuring one-way heat transport and immunity to backscattering. In contrast, classical nonreciprocal systems apply nonzero Weyl node separation only at the boundary (right panel of Fig. 3a). This localized modulation leads to discrepancies in energy flow between the outer and inner boundary regions, resulting in nonreciprocal heat transfer (Fig. 3b). However, this nonreciprocal heat flow is fragile due to mismatches in material properties and radiation orientations between the bulk and boundary, as evidenced by the projective band structures (Fig. 3c). In classical nonreciprocal systems, this mismatch shifts in-gap chiral edge states out of the bandgap, resulting in local unidirectionality but lacking global topological protection.

To quantitatively evaluate the nonreciprocity, we calculate the spectral heat transfer for both the topological nonreciprocal and nonreciprocal lattices in the same hexagonal geometry (Fig. 4d). The frequency is set within the first nontrivial bandgap ($\omega = 1280$ cm$^{-1}$). An isotropic emitter, maintained at 350 K, is placed at one corner of the hexagon, while a receiver at 300 K is positioned at an adjacent corner. The net power received by particle $j$ can be expressed as [29]:

$$\langle \mathcal{P}_{\omega,j} \rangle = \sum_{i \neq j} (\mathcal{P}_{\omega,ij} - \mathcal{P}_{\omega,ji})$$
$$= \sum_{k \neq j} \frac{d\omega}{2\pi} [\Theta(\omega, T_i)\mathcal{T}_{ji}(\omega) - \Theta(\omega, T_j)\mathcal{T}_{ij}(\omega)] \quad (3)$$

where $\mathcal{T}$ is the transmission coefficient, with calculation details provided in Supplementary Note II. In the topological nonreciprocal system, forward energy transfer $\mathcal{P}_{\omega,ij}$ is significantly stronger than backward transfer $\mathcal{P}_{\omega,ji}$, especially as the Weyl node separation $b$ approaches 0.8 nm$^{-1}$. Although the classical nonreciprocal lattice exhibits a similar behavior, the nonreciprocity and the energy transfer intensity are substantially weaker than that in the nontrivial lattice. In Fig. 4e, we fix the Weyl node separation at $b = 0.8$ nm$^{-1}$ to analyze the nonreciprocal response across a broad frequency range. The topological nonreciprocal system shows a significant advantage in heat transfer intensity, particularly near the first nontrivial band gap. Furthermore, by tuning the Weyl node separation, the topological nonreciprocal system achieves over four orders of magnitude enhancement in heat transfer, as illustrated in Fig. 4f.

*Robust one-way heat transport.* Robustness against disorder and defects is a pivotal advantage of topological systems, offering substantial potential for practical applications by relaxing strict fabrication constraints. To assess this robustness, we introduce artificial scattering obstacles, represented by white circles in Figs. 4b and 4e. In the topological nonreciprocal system, energy flow consistently follows the edge of the nontrivial lattice, while in the classical nonreciprocal system, energy flow is obstructed and exhibits pronounced backscattering. We further introduce disorder by randomly displacing particles within a quarter of the lattice constant (Figs. 4e and 4f). Despite this disorder, the topological nonreciprocal lattice maintains strong one-way heat transport, surpassing the performance of the classical nonreciprocal system. Although the underlying physics in deep-subwavelength systems is more intricate, we provide a straightforward proof using tight-binding theory (see Supplementary Note IV). Wave packet dynamics further illustrate that topologically nonreciprocal lattices exhibit significant advantages over classical nonreciprocal systems, especially in terms of robustness and transmission efficiency.

**Conclusion**

This work reveals robust topological one-way radiative heat transfer in Weyl-semimetal-based thermophotonic lattices, showcasing exceptional robustness against obstacles and disorder. In comparison to trivial lattices, the radiative heat flux in nontrivial lattices is enhanced by several orders of magnitude and offers tunability. The underlying mechanism is attributed to nontrivial band topology and many-body long-range interactions. These findings pave the way towards energy harvesting and manipulating thermal radiation, and immediately stimulate new thoughts for topological thermal physics and thermophotonic applications.

**Acknowledgements**

C.-W.Q. acknowledged the financial support by the Ministry of Education, Republic of Singapore (Grant No.: A-8000107-01-00), the National Research Foundation, Singapore (NRF) under NRF's Medium Sized Centre: Singapore Hybrid-Integrated Next-Generation $\mu$-Electronics (SHINE) Centre funding programme.

**Competing interests**

The authors declare no competing interests.

**References**

[1] A. Fresnel, Theory of light. Fifth section: various questions of optics, *Ann. Chim. Phys.* 9, 57 (1818).

[2] H. Fizeau, On the relative hypotheses, luminous ether, and on an experiment which seems to demonstrate that the movement of bodies changes the speed with which light propagates in their interior, *CR Hebd. Acad. Sci.* **33**, 349 (1851).

[3] L. J. Xu, G. Q. Xu, J. P. Huang, C. W. Qiu, Diffusive Fizeau Drag in Spatiotemporal Thermal Metamaterials, *Phys. Rev. Lett.* **128**, 145901 (2022).

[4] P. A. Huidobro, E. Galiffi, S. Guenneau, R. V. Craster, and J. B. Pendry, Fresnel drag in space–time-modulated metamaterials, *Proc. Natl. Acad. Sci. U.S.A.* **116**, 24943 (2019).

[5] P.-C. Kuan, C. Huang, W. S. Chan, S. Kosen, and S.-Y. Lan, Large Fizeau's light-dragging effect in a moving electromagnetically induced transparent medium, *Nat. Commun.* **7**, 13030 (2016).

[6] Y. Dong, L. Xiong, I. Y. Phinney, Z. Sun, R. Jing, A. S. McLeod, S. Zhang, S. Liu, F. L. Ruta, H. Gao, Z. Dong, R. Pan, J. H. Edgar, P. Jarillo-Herrero, L. S. Levitov, A. J. Millis, M. M. Fogler, D. A. Bandurin, and D. N. Basov, Fizeau drag in graphene plasmonics. *Nature* **594**, 513-516 (2021).

[7] W. Y. Zhao, S. H. Zhao, H. Y. Li, S. Wang, S. X. Wang, M. I. B. Utama, S. Kahn, Y. Jiang, X. Xiao, S Yoo, K. Watanabe, T. Taniguchi, A. Zettl, and F. Wang, Efficient Fizeau drag from Dirac electrons in monolayer graphene, *Nature* **594**, 517–521 (2021).

[8] S. A. H. Gangaraj, B. Y. Jin, C. Argyropoulos, and F. Monticone, Enhanced Nonlinear Optical Effects in Drift-Biased Nonreciprocal Graphene Plasmonics, *ACS Photonics* **10**, 3858-3865 (2023).

[9] C. L. Zhou, L. Qu, Y. Zhang, and H. L. Yi, Enhancement and active mediation of near-field radiative heat transfer through multiple nonreciprocal graphene surface plasmons, *Phys. Rev. B* **102**, 245421 (2020).

[10] Y. Zhang, C. L. Zhou, H. L. Yi, H. P. Tan, Radiative Thermal Diode Mediated by Nonreciprocal Graphene Plasmon Waveguides, *Phys. Rev. Applied* **13**, 034021 (2020).

[11] F. D. M. Haldane, Model for a Quantum Hall Effect without Landau Levels: Condensed-Matter Realization of the "Parity Anomaly", *Phys. Rev. Lett.* **61**, 2015 (1988).

[12] Y. Hatsugai, Chern number and edge states in the integer quantum Hall effect, *Phys. Rev. Lett.* **71**, 3697–3700 (1993).

[13] X. L. Qi, and S. C. Zhang, Topological insulators and superconductors, *Rev. Mod. Phys.* **83**, 1057–1110 (2011).

[14] B. Bahari, A. Ndao, F. Vallini, A. E. Amili, Y. Fainman, and B. Kanté, Nonreciprocal lasing in topological cavities of arbitrary geometries, *Science* **358**, 636-640 (2017).

[15] W. X. Zhang, F. X. Di, X. G. Zheng, H. J. Sun, and X. D. Zhang, Hyperbolic band topology with non-trivial second Chern numbers, *Nat. Commun.* **14**, 1083 (2023)

[16] S. R. Pocock, X. F. Xiao, P. A. Huidobro, and V. Giannini, Topological Plasmonic Chain with Retardation and Radiative Effects, *ACS Photonics* **5**, 2271-2279 (2018).

[17] A. Ott, Z. H. An, A. Kittel, and S.-A. Biehs, Thermal near-field energy density and local density of states in topological one-dimensional Su-Schrieffer-Heeger chains and two-dimensional Su-Schrieffer-Heeger lattices of plasmonic nanoparticles, *Phys. Rev. B* **104**, 165407 (2021).

[18] B. X. Wang, and C. Y. Zhao, Topological phonon polariton enhanced radiative heat transfer in bichromatic nanoparticle arrays mimicking Aubry-André-Harper model, *Phys. Rev. B* **107**, 125409 (2023).

[19] S. Guddala, F. Komissarenko, S. Kiriushechkina, A. Vakulenko, M. Li, V. M. Menon, A. Alù, and A. B. Khanikaev, Topological phonon-polariton funneling in midinfrared metasurfaces, *Science* **374**, 225-227 (2021).

[20] S. H. Yang, G. Q. Xu, X. Zhou, J. X. Li, X. H. Kong, C. L. Zhou, H. Y. Fan, J. F. Chen, and C.-W. Qiu, Hierarchical bound L. states in heat transport, *Proc. Natl. Acad. Sci. U.S.A.* **121**, e2412031121 (2024).

[21] J. D. H. Rivero, C. Fleming, B. K. Qi, L. Feng, and L. Ge, Robust Zero Modes in Non-Hermitian Systems without Global Symmetries, *Phys. Rev. Lett.* **131**, 223801 (2023).

[22] L. X. Zhu, and S. H. Fan, Persistent Directional Current at Equilibrium in Nonreciprocal Many-Body Near Field Electromagnetic Heat Transfer, *Phys. Rev. Lett.* **117**, 134303 (2016).

[23] P. Ben-Abdallah, Photon Thermal Hall Effect, *Phys. Rev. Lett.* **116**, 084301 (2016).

[24] C. Guo, V. S. Asadchy, B. Zhao, and S. H. Fan, Light control with Weyl semimetals, *elight* **3**, 2 (2023).

[25] B. Zhao, C. Guo, C. A. C. Garcia, P. Narang, and S. H. Fan, Axion-Field-Enabled Nonreciprocal Thermal Radiation in Weyl Semimetals, *Nano Lett.* **20**: 1923-1927 (2020).

[26] S. Nie, T. Hashimoto, and F. B. Prinz, Magnetic Weyl semimetal in $K_2Mn_3(AsO_4)_3$ with the minimum number of Weyl points, *Phys. Rev. Lett.* **128**, 176401 (2022).

[27] J.-R. Soh, F. de Juan, M. G. Vergniory, N. B. M. Schröter, M. C. Rahn, D. Y. Yan, J. Jiang, M. Bristow, P. A. Reiss, J. N. Blandy, Y. F. Guo, Y. G. Shi, T. K. Kim, A. McCollam, S. H. Simon, Y. Chen, A. I. Coldea, and A. T. Boothroyd, Ideal Weyl semimetal induced by magnetic exchange, *Phys. Rev. B* **100**, 201102 (2019).

[28] P. Ben-Abdallah, and S.-A. Biehs, Many-Body Radiative Heat Transfer Theory, *Phys. Rev. Lett.* **107**, 114301 (2011).

[29] S. R. Pocock, X. F. Xiao, P. A. Huidobro, and V. Giannini, Topological Plasmonic Chain with Retardation and Radiative Effects, *ACS Photonics* **5**, 2271-2279 (2018).

[30] S.-A. Biehs, R. Messina, P. S. Venkataram, A. W. Rodriguez, J. C. Cuevas, and P. Ben-Abdallah, Near-field radiative heat transfer in many-body systems, *Rev. Mod. Phys.* **93**, 025009 (2021).

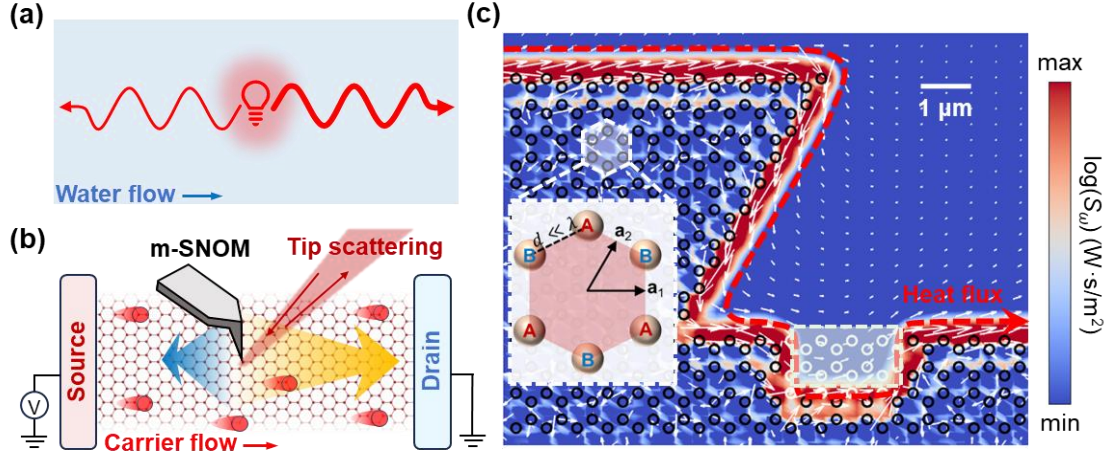

**FIG. 1. The impact of external momentum on wave/energy transmission mechanisms.**
(a) Schematic of the Fresnel light dragging effect. The blue arrow indicates the water flow direction, while the red arrows, varying in thickness, represent the different speeds of forward and backward light propagation [*Ann. Chim. Phys.* 9, 57 (1818); *C. R. Acad. Sci.* 33, 349 (1851)]. (b) Schematic of the plasmonic Fizeau effect, with bias voltage applied at the source/drain terminals. Red spheres represent drifting Dirac electrons, while yellow and blue arrows denote forward and backward nonreciprocal surface plasmon polaritons. Near-field excitation and detection are performed using a cryogenic magneto scanning near-field optical microscope (m-SNOM) [*Nature* 594, 513-516 (2021)]. (c) Illustration of robust one-way heat flux in a Weyl-semimetal-based thermophotonic crystal. The light green shaded area in the bottom right indicates non-magnetic scattering defects. The inset on the left shows the unicell of the thermophotonic lattice, with lattice vectors $\mathbf{a}_1$ and $\mathbf{a}_2$ determining the periodicity of the thermal lattice.

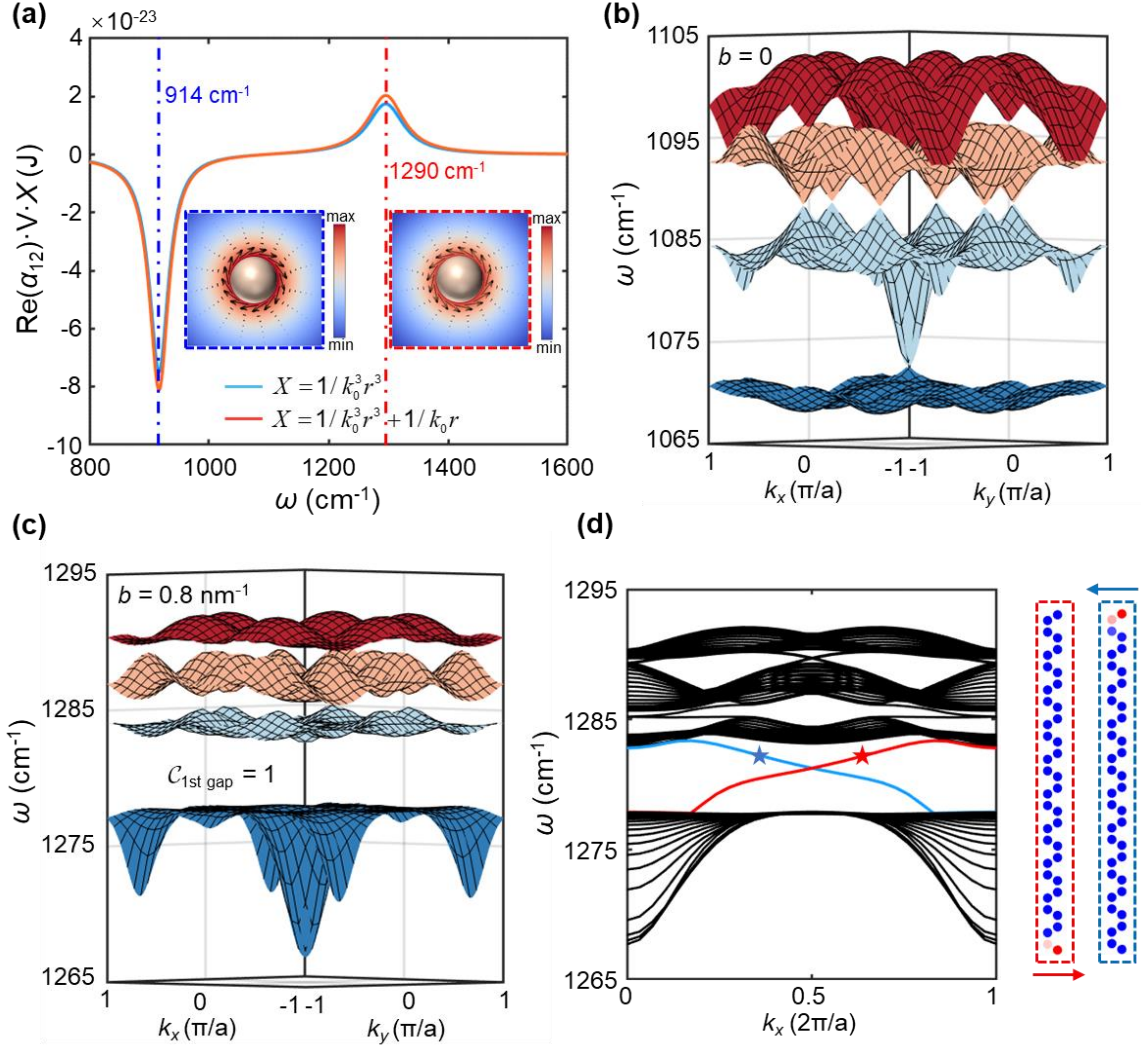

**FIG. 2. Circular heat flux and thermophotonic band structures.**

(a) Electronic contribution to $\alpha'_{12}\Theta(\omega, T_p)k_0^3$ at $T_p = 300$ K, weighted by $1/(k_0 r)^3$ (near-field regime) and $X = 1/(k_0 r)$ (far-field regime). The insets depict the in-plane circular heat flux of a single nanoparticle. (b) Bulk band structure of a trivial thermophotonic lattice, where all four bands degenerate at high-symmetry points. (c) Bulk band structure of a nontrivial thermophotonic lattice, with all four bands fully separated. (d) Projective band structure of the 1D supercell, featuring a nontrivial bandgap with Weyl nodes separated by $b = 0.8$ nm$^{-1}$. The red and blue solid lines indicate gapless chiral one-way edge states, with the right panel depicting the chiral edge modes corresponding to the red and blue stars in the projective band structure.

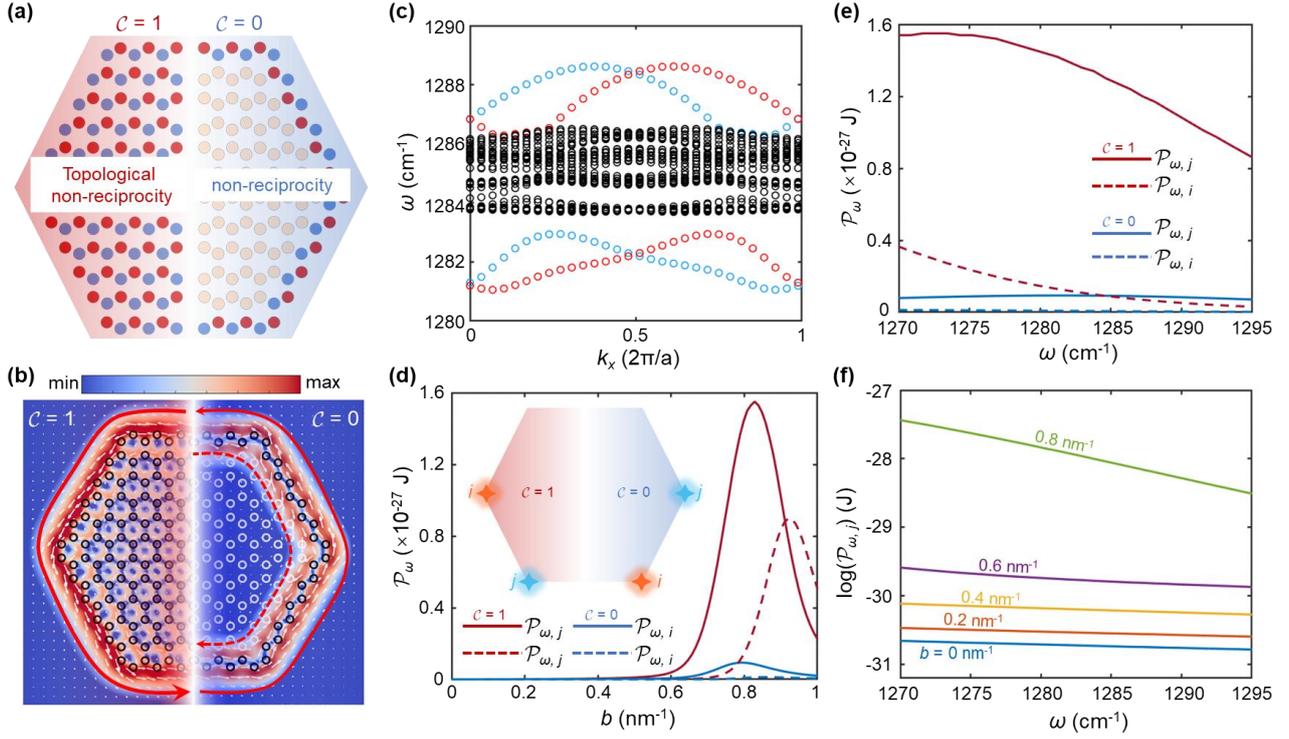

**FIG. 3. Topological nonreciprocity vs. classical nonreciprocity.**

(a) Schematic of topological and classical nonreciprocal systems, with red and blue circles representing nanoparticles with nonzero Weyl momentum node separation. (b) In-plane heat flux in topological nonreciprocal (left) and classical nonreciprocal (right) thermophotonic lattices. Black circles denote nanoparticles with Weyl node separation $b = 0.8$ nm$^{-1}$, while white circles represent nanoparticles with $b = 0$ nm$^{-1}$. (c) Projective band structure of a classical nonreciprocal supercell. (d) Spectral energy transfer as a function of Weyl node separation distance for both topological and classical nonreciprocal systems, with the frequency fixed at $\omega = 1280$ cm$^{-1}$. (e) Comparison of spectral energy transfer between the topological and classical nonreciprocal systems at a Weyl node separation distance of $b = 0.8$ nm$^{-1}$. (f) Spectral energy transfer in the topological nonreciprocal system across varying Weyl node separations.

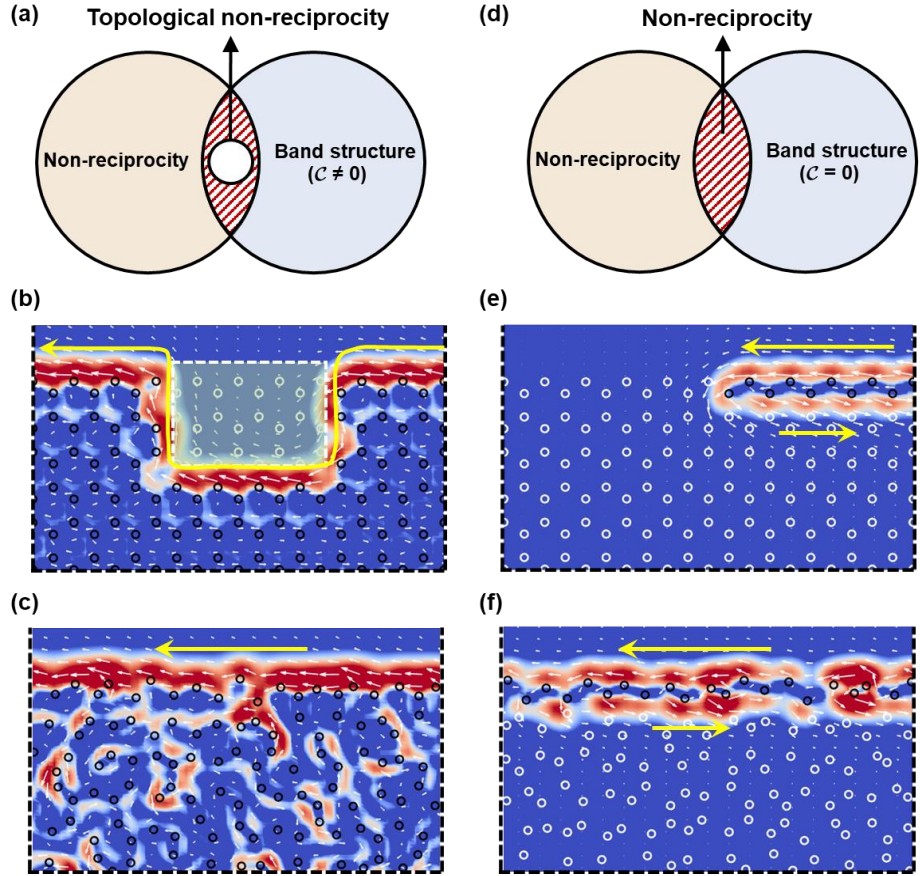

**FIG. 4. Robustness of thermophotonic one-way edge states.**

(a) The interplay between nonreciprocal thermophotonics and topological physics. (b) In-plane heat flux in nontrivial thermophotonic lattices with non-magnetic scatterer defects. (c) In-plane heat flux in nontrivial thermophotonic lattices with positional disorder, where disorder corresponds to particle displacements within one-quarter of the lattice constant. (d) Schematic of a trivial nonreciprocal system with nonreciprocal behavior but trivial bulk topology. (e) and (f) depict in-plane heat flux for classical reciprocal systems under the same configurations as (b) and (c). Black circles represent nontrivial nanoparticles, while white circles represent trivial nanoparticles.